\documentclass[a4paper,fleqn,usenatbib]{mnras}

\usepackage[T1]{fontenc}
\usepackage{ae,aecompl}
\usepackage{bm}

\usepackage{graphicx}   
\usepackage{amsmath}    
\usepackage{amssymb}    
\usepackage{amsfonts}
\usepackage{verbatim}
\usepackage{mathrsfs}   
\usepackage{setspace}
\usepackage{mathrsfs}
\usepackage{bm}
\usepackage{verbatim}
\usepackage{xcolor}
\usepackage[multiple]{footmisc}

\usepackage{lipsum}

\title[Luminosity of highly magnetic massive white dwarfs]{Suppression of luminosity and mass--radius relation of highly magnetized white dwarfs}

\author[Gupta, Mukhopadhyay \& Tout]{
	Abhay Gupta,$^{1}$\thanks{gabhay@iisc.ac.in}
	Banibrata Mukhopadhyay$^{1}$\thanks{bm@iisc.ac.in} and
	Christopher A. Tout$^2$\thanks{cat@ast.cam.ac.uk}
	\\
	$^{1}$ Department of Physics, Indian Institute of Science, Bangalore 560012, India\\
	$^2$ Institute of Astronomy, The Observatories, Medingley Road, Cambridge CB3 OHA, UK
}

\date{Accepted 2020 June 2. Received 2020 April 22; in original form 2020 February 14}

\pubyear{2020}

\begin{document}
	
	\label{firstpage}
	\pagerange{\pageref{firstpage}--\pageref{lastpage}}
	\maketitle
	
	\begin{abstract}
		We explore the luminosity $L$ of magnetized white dwarfs and its effect on the mass--radius relation.
		We self-consistently obtain the interface between the electron degenerate gas
		dominated inner core and the outer ideal gas surface layer or envelope by incorporating both the components of gas
		throughout the model white dwarf. This is obtained by solving the set of magnetostatic equilibrium, photon diffusion 
		and mass conservation equations in the Newtonian framework, for different sets of luminosity and magnetic field. 
		We appropriately use magnetic opacity, instead of 
		Kramer's opacity, wherever required. 
		We show that the Chandrasekhar-limit 
		is retained, even at high luminosity upto about $10^{-2}\,L_\odot$ but without
		magnetic field, if the temperature is set constant 
		inside the interface. However there is an increased mass for large-radius white dwarfs, an effect of photon diffusion.
		Nevertheless, in the presence of strong magnetic fields, with central strength
		of about $10^{14}$\,G, super-Chandrasekhar 
		white dwarfs, with masses of about $1.9\,M_{\odot}$, are obtained even when the temperature inside the interface is kept 
		constant. Most interestingly, small-radius magnetic white dwarfs remain super-Chandrasekhar even if their luminosity 
		decreases to as low as about $10^{-20}\,L_{\odot}$. However, their large-radius counterparts in the same mass--radius
		relation merge with Chandrasekhar's result at low $L$. Hence, we argue for the possibility of highly magnetized, low
		luminous super-Chandrasekhar mass white dwarfs which, owing to their faintness, can be practically hidden. 
	\end{abstract}
	
	
	
	
	\section{Introduction}
	
	Observations of several overluminous type Ia supernovae (see, \citealt{Howell,Scalzo,Silver})
	imply the existence of super-Chandrasekhar white dwarfs. It was shown earlier (\citealt{Das-MukhoPRD, 
		Das-MukhoPRL, Satto}) that white dwarfs with highly super-Chandrasekhar mass, $M > M_{\rm Ch} = 1.44\,M_{\odot}$, 
	are possible when rotation and magnetic fields are taken into consideration. 
	It was also shown (\citealt{Das-MukhoPRD,Das-MukhoPRL}) that when the magnetic field is greater than a critical strength
	of $4.414 \times10^{13}$\,G, such that the Larmor radius is same order of or smaller than the Compton wavelength of the 
	electrons, the effect of Landau quantization becomes important and thus modifies the equation of state
	(EoS) of the electron degenerate matter of the white dwarf. This modified EoS further gives 
	rise to super-Chandrasekhar white dwarfs and a new limit of $2.58\,M_\odot$. Interestingly, 
	Sloan Digital Sky Survey (SDSS) data (\citealt{SDSS}) indicate that magnetic 
	samples of white dwarfs span the same temperature range as non-magnetic white dwarfs and 
	support the claim that magnetic white dwarfs tend to have larger masses compared to their non-magnetic 
	counterparts.
	
	The existence of magnetic white dwarfs (hereinafter B-WDs) does not just explain super-Chandrasekhar mass 
	progenitors of overluminous type Ia supernovae but also (\citealt{Rao}) 
	B-WDs correspond to soft gamma-ray repeaters (SGRs) and anomalous
	X-ray pulsars (AXPs) with much lower magnetic fields than magnetars. 
	Further, the glitch and anti-glitch of the source 1E 2259+586 may be explained based on the stiffness of 
	the magnetic field profile, mass loss and rotational energy extraction rates of a B-WD. Moreover, the 
	peculiar radio transient of GCRT J1745--3009 can be interpreted as a white dwarf pulsar (WDP) based on the
	B-WD model, because its larger magnetic field and smaller radius could prove it is of low luminosity and further away and hence dimmer and able to evade detection.
	
	However, the questions remain, why have not enough B-WDs yet been seen directly and why have only about a dozen overluminous
	type Ia supernovae implying super-Chandrasekhar white dwarfs have been seen so far? While \cite{DMR13} attempted
	to address these questions by arguing that they are rare objects as overluminous type Ia supernovae, 
	\cite{Mukul} explored their luminosity with the model described by \cite{Shapiro}
	for nonmagnetic white dwarfs. Based on the photon diffusion equation and the magnetostatic balance condition, and
	assuming that interface parameters remain unaffected between B-WDs and nonmagnetic white dwarfs, 
	\cite{Mukul} found that, for a given age, the
	luminosity of B-WDs decreases significantly, from about $10^{-6}$ to $10^{-9}\,L_{\odot}$ when the magnetic 
	field strength increases from about $10^{9}$ to $10^{12}\,$G at the interface and hence the envelope. 
In fact, \cite{2014Natur.515...88V}, based on optical observation, showed that the strong magnetic field suppresses convection
over the entire surface in magnetic white dwarfs, which decreases luminosity and increases cooling time scale significantly. 
They analysed about eight years observed data and obtained brightness and magnetic 
field of WD~1953-011 which is a cool white dwarf. Their results argue that the magnetic fields are expected to be common 
in cool, hence low luminous, white dwarfs and B-WDs are actually too young and low luminous. Indeed, in the theoretical ground,
\cite{Mukul} found the cooling rates corresponding to luminosities from about $10^{-6}$ to $10^{-9}\,L_{\odot}$.
	In this way, they studied the effect of luminosity on the outer surface layer of the white dwarf, 
	where only ideal gas pressure exists. Because the interface between the electron degenerate core and the surface 
	layer or envelope is very close to the surface of the white dwarf, the mass at the interface was chosen to be the 
	total mass of the corresponding white dwarf. Thus, the characteristics of the white dwarf were
	obtained by solving the two differential equations of magnetohydrostatic equilibrium and photon diffusion. 
	While this was a very good 
	start, the choice of the same interface parameters for B-WDs and nonmagnetic white dwarfs
	is a big assumption. 
	Moreover, the results based on modelling only the envelope with preassigned mass and radius should 
	also be taken with caution, because B-WD masses and luminosities are yet be established.
	
	Here we remove all the above assumptions and find the mass--radius relation along with
	luminosity of B-WDs self-consistently. 
Earlier, the mass--radius relation was rigorously explored in Newtonian (\citealt{Das-MukhoPRD,DMR13}) as well as general 
relativistic (\citealt{UD14,Satto,UD15,mrb17}) formalisms in various magnetic field and rotational configurations, by our group. Some of these works
considered poloidally as well as toroidally dominated magnetic field geometries and obtained similar results showing existence of super-Chandrasekhar
white dwarfs (also see, e.g., \citealt{ruffini,franzon,2018EPJC...78..411C}). Recently \cite{Otoniel_2019} revisited the problem again rigorously in general relativistic frameworks with dipole fields, 
but more importantly considering possible instability of the matter with respect to pycnonuclear and electron capture reactions.
Such stability criteria, particularly of pycnonuclear reactions, were not considered by our group in order to establish mass--radius relation mentioned above. 
However, still \cite{Otoniel_2019} found the maximum mass of nonrotating B-WDs to be $2.14M_\odot$ with central magnetic 
field $\sim 3.85\times 10^{14}$ G with limiting central density $9.35\times 10^9$ g cm$^{-3}$, when rotation certainly
would increase the mass further. According to them, above this 
density,  pycnonuclear fusion reactions destabilize the star. This work further establishes strongly the idea of the existence
of significantly super-Chandrasekhar white dwarf with a new mass-limit, what we initiated in 2012 with a simpler model and 
evolved over the years with the introduction of rigorous physics. Nevertheless, all the work neglected any possible variation
of temperature in the star and its effect in the mass--radius relations. However, this is very important when luminosity of B-WDs is 
also planned to be explored and the aim is to understand why super-Chandrasekhar white dwarfs are hardly visible. \cite{Mukul} included
temperature gradient in assessing luminosity of B-WDs, but they did not explore mass--radius relation and confined their
physics in the envelope strictly. Theirs was just one step forward compared to white dwarf luminosity investigated 
by \cite{Shapiro} without magnetic fields. In the present work we have removed such limitations.

We model the star throughout, from the core to surface, with the finite 
	temperature gradient in envelope determining the underlying luminosity with total pressure as
	the sum of the electron degenerate, ideal 
	gas and magnetic pressures. Naturally the correct modelling leads to the dominance of electron degenerate
	pressure over ideal gas pressure in core and otherwise in the envelope. The radius where the ideal gas pressure
	dominates the degenerate pressure is defined to be the interface. We begin 
	to solve the three differential equations for
	magnetostatic equilibrium, mass conservation and photon diffusion. 
As we mainly venture (strongly) magnetized white dwarfs, we neglect
any possible convection as strong field suppresses convection effects
(e.g. \citealt{1992ApJ...389..724C,2003A&ARv..11..153S}). Moreover, we use magnetic 
	opacity instead of Kramer's opacity, wherever necessary. 
	
	In the next section we give an overview of the procedure 
	to self-consistently obtain the interface of the B-WD. In section 3, we describe the mass--radius 
	relation including the effects of luminosity and, hence, finite temperature gradient, and magnetic fields. 
	Subsequently, in section 4 we attempt to reinterpret the results based on total energy conservation.  
	In section 5 we summarize the results and in section 6 conclude with some discussion.  
	
	
	
	\section{White dwarf and its interface}
	
	We describe the conditions used and equations solved to formulate a method to self-consistently 
	obtain the structure of B-WDs and the interface between the inner electron degenerate core and outer ideal 
	gas envelope. For the present purpose, we neglect any 
general relativistic effect as it is not expected to influence luminosity,
which mostly is the envelope phenomena. Moreover, as the main plan is to target
significantly super-Chandrasekhar white dwarfs, possible slight change in mass
using general relativistic framework compared to their Newtonian counterpart does not matter. Therefore, the model equations describing magnetostatic equilibrium, photon diffusion and mass conservation, approximating stars to be spherical
without losing any physics for the present purpose, 
	respectively are 
	\begin{equation}
	\frac{\mathrm{d}}{\mathrm{d} r}\left(P_{\rm degenerate}+P_{B}+P_{\rm ideal}\right)=-\frac{G m(r)}{r^{2}}\left(\rho+\rho_{B}\right),
	\label{eq:magnetostatic}
	\end{equation}
	\begin{equation}
	\frac{\mathrm{d} T}{\mathrm{d} r}=-\frac{3}{4 a c} \frac{\kappa\left(\rho+\rho_{B}\right)}{T^{3}} \frac{L_r}{4 \pi r^{2}},
	\label{eq:photon diffusion}
	\end{equation}
	\begin{equation}
	\frac{d m}{d r}=4 \pi r^{2} (\rho+\rho_{B}).
	\label{eq:mass conservation}
	\end{equation}
	In these equations, $P_{\rm degenerate}$ and $P_{\rm ideal}$ are the electron degeneracy pressure 
	established by \cite{Chandra} and the ideal gas pressure ($\rho kT/\mu m_{\rm p}$, where $\rho$ is
	the density of the matter, $k$ Boltzmann's constant, $T$ the temperature, $\mu \approx 2$ the mean 
	molecular weight and $m_{\rm p}$ the mass of proton) respectively, $m$ is the mass
	enclosed at radius $r$, 
	$G$ the Newton's Gravitation constant, $a$ the 
	radiation constant, $c$ the speed of light, $L_r$ the luminosity at radius $r$, $\kappa$ the opacity,
	$P_{B}$ and $\rho_{B}$ the contributions from magnetic field to the pressure and density respectively
	and $r$ is the distance from the centre of star. As we restrict central
magnetic field to $\sim 10^{14}$ G, the assumption of spherical star is 
approximately justified (\citealt{Satto}).
	
	It is well-known that compact stars with strong magnetic fields (e.g. magnetars, B-WDs) tend to have 
	additional pressure and density contributions owing to the magnetic field which can be denoted by $P_B=B^{2}/8\pi$ 
	where field $B = \sqrt{\textbf{\emph{B.B}}}$ and $\rho_B=B^{2}/8\pi c^2$ (\citealt{Monika,UD14,Mukul}). 
	The opacity for a non-magnetized white dwarf is approximated by Kramers' formula, 
	$ \kappa = \kappa_{0}\rho T^{-3.5}$, where $\kappa_{0}=4.34 \times 10^{24} Z(1+X) \mathrm{cm}^{2} \mathrm{g}^{-1}$ and $X$ and $Z$ are the mass fractions of hydrogen and heavy elements (elements other than hydrogen and helium) 
	in the stellar interior. For a typical white dwarf, $X$ = 0, and we assume for simplicity the mass fraction of 
	helium $Y = 0.9$ and $Z = 0.1$. The opacity is due to the bound$-$free and free$-$free transitions of 
	electrons (\citealt{Shapiro}). In the presence of $B$, the variation of radiative 
	opacity with $B$ can be modelled similarly to neutron stars as $\kappa = \kappa_{B} \approx 5.5 \times 10^{31} \rho T^{-1.5} B^{-2} \mathrm{cm}^{2} \mathrm{g}^{-1}$ (\citealt{Ventura}), with the condition that radiative opacity should be dominant and $B/10^{12}$G $\geq$ $T/10^{6}$K, otherwise Kramers' formula should be used.
	For the profile of magnetic field inside the B-WD we use the profile used earlier extensively for 
	magnetized neutron stars and B-WDs (\citealt{Bando,UD14}) given by 
	\begin{equation}
	B\left(\frac{\rho}{\rho_{0}}\right)=B_{\rm s}+B_{0}\left[1-\exp \left(-\eta\left(\frac{\rho}{\rho_{0}}\right)^{\gamma}\right)\right],
	\label{eq: magnetic}
	\end{equation}
	where $\rho_{0}$ is a measure of density (chosen roughly $10$ per cent of the central matter density of the 
	corresponding white dwarf), $B_{\rm s}$ is the surface magnetic field, $B_{0}$ is a parameter with dimension of 
	magnetic field and $\eta$ and $\gamma$ are dimensionless parameters which determine how exactly the field decays 
	from the centre to surface. This can be parameterized (\citealt{UD14}) with suitable constraints 
	on pressure. Note that, as $\rho \rightarrow 0$ close to the surface of the white dwarf, $B \rightarrow B_{\rm s}$. 
	For our work, the parameters are selected to be $\rho_{0} = 10^{9}$~g~cm$^{-3}$, $\eta=0.8$ and $\gamma=0.9$ 
	following \cite{Mukul}. Also, $B_{\rm s}$ and $B_{0}$ are chosen based on previous 
	work and observations (\citealt{UD14,Mukul}).
	
	In order to obtain a solution, luminosity is taken to be constant for simplicity and because there is no hydrogen burning or nuclear fusion taking place to power the white dwarf, so that $L_r=L$. 
Note that in principle convection may transfer a significant fraction 
of the total energy flux depending on magnetic field strength. It is 
however also known (e.g. \citealt{1992ApJ...389..724C,2003A&ARv..11..153S}) that
strong field significantly inhibits convection. As the present venture is mostly to analyse strong field effects in white dwarfs, when even the weakest 
chosen surface field is no less than $10^7$ G, we can safely ignore 
convection from the model equations.
	\cite{Koester} argued that, if the luminosity is considered to be constant, it is preferred to 
	integrate the differential equations from the surface towards the centre. The surface is defined by the 
	corresponding radius of the white dwarf. 
	In order to solve for the set of differential equations, here by a Runge-Kutta method, the 
	different parameters such as temperature, mass and density at the surface must be supplied.  
	The surface temperature is obtained from the Stefan-Boltzmann law, 
	$T_{\mathrm{s}}=(L / 4 \pi R^{2} \sigma)^{1 / 4}$, for different luminosities 
	$L$ and the corresponding 
	radii $R$, when $\sigma$ is the Stefan-Boltzmann constant. The surface density, which has to be several orders of magnitude smaller than the central density, 
	is chosen such that the equations can be solved simultaneously within the tolerance limits, 
	because both 1~g~cm$^{-3}$ and $10^{-10}$~g~cm$^{-3}$ can be considered to be
	quite small compared to $10^4$ to $10^9$~g~cm$^{-3}$ 
	(the range of central densities for white dwarfs). The optimum range comes out to be 
	$10^{-6}$ to $10^{-9}$~g~cm$^{-3}$ and we choose $10^{-7}$~g~cm$^{-3}$ as the surface density for all the cases. 
	We have checked that the results are unaffected by any minor change in the 
	surface density within the above range. Further, the mass is obtained with a shooting method 
	by iteratively solving the equation set to satisfy the condition that the mass 
	obtained close to the centre, at $10$ km from the centre, integrating from surface is same as 
	that obtained by solving mass conservation equation 
	(\ref{eq:mass conservation}) given the solution for density profile. 
	In this way we obtain the mass for a given radius of a white dwarf.
	
	Now we can self-consistently obtain the interface, which is defined to be the radius at which the ideal gas pressure
	starts to dominate over the degenerate pressure. An example has been discussed in the Appendix to further clarify
	how the interface of a white dwarf is obtained in our method. It is known that the interior of a white dwarf 
	consists completely of degenerate gas. The electrons have a large mean free path because of the filled Fermi 
	sea, so there is a high thermal conductivity and hence a uniform temperature (\citealt{Shapiro}). 
	Thus, another condition can be imposed that the R.H.S. of photon-diffusion equation (\ref{eq:photon diffusion}) is 
	zero from the interface to centre and we study its effects on the mass--radius solution.
	We obtain interesting results on implementing this condition in non-magnetized white dwarfs. 
	Hence, there are two possible solution cases, (1) the photon-diffusion equation (\ref{eq:photon diffusion}) 
	is valid throughout the white dwarf, and (2) $dT/dr=0$ for radii less than the interface radius. 
	However, both cases produce a very similar interface radius.
	
	\subsection{Nonmagnetic results}
\begin{figure}
		\includegraphics[width=\columnwidth]{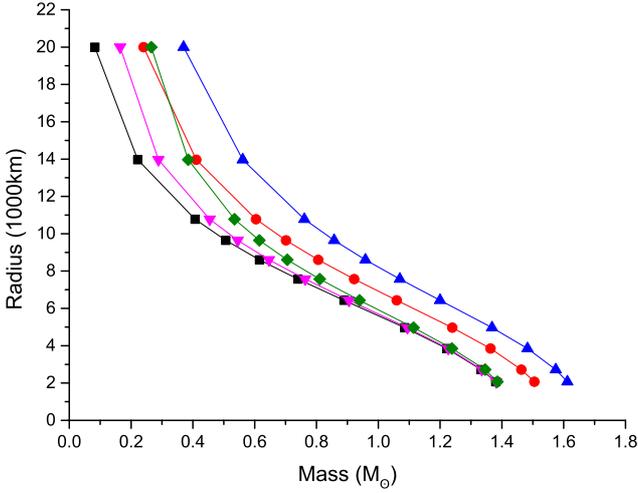}
		\caption{The effect of radiative energy transport to the mass--radius relation, for constant
			temperature throughout, Chandrasekhar's result (black squares), non-zero $\frac{dT}{dr}$ 
			throughout for $L=10^{-2}\,L_{\odot}$ (blue upward triangles) and $10^{-4}\,L_{\odot}$ (red circles), and $\frac{dT}{dr}=0$  
			below the interface radius for $10^{-2}\,L_{\odot}$ (green diamonds) and $10^{-4}\,L_{\odot}$ (magenta downward triangles).}
		\label{fig:1}
	\end{figure}

	First we investigate the effect of a finite temperature gradient in the stellar structure without magnetic fields.
	Fig. \ref{fig:1} shows how the photon diffusion equation, the temperature gradient, modifies the mass--radius 
	relation compared to Chandrasekhar's original results (\citealt{Chandra}), when $dT/dr\neq 0$ throughout, the first condition above. 
	We find that there is a shift in the mass--radius relation towards increasing mass and radius compared to the case with $dT/dr=0$ throughout. It shows super-Chandrasekhar masses for smaller radii. 
	However, there seems to be enough evidence that, in the absence of magnetic fields and rotation, the Chandrasekhar-limit is
	preserved.
	By imposing the second condition above, we see that indeed the Chandrasekhar limit is retained but there is an overall 
	increase in mass in the larger radius regime in the mass--radius relation compared to Chandrasekhar's. 
	This is similar to the earlier finding of increasing mass seen at larger radii, while preserving Chandrasekhar's limit, 
	at higher temperatures with the finite temperature electron degenerate EoS (see fig 4. of 
	\citealt{Rueda}) and the same is confirmed by SDSS. Therefore, the first condition above seems
	physically implausible. Note that in either of the cases, increasing luminosity leads to more deviation
	of the mass--radius relation with increasing mass (and radius for the first case). This is expected because
	higher luminosity indicates higher thermal energy leading to higher ideal gas pressure, allowing the white dwarf to 
	hold more mass.
	
	\begin{figure}
		\includegraphics[width=\columnwidth]{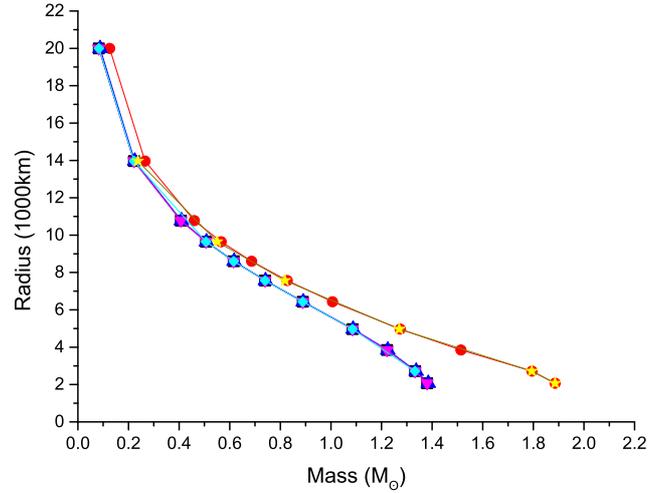}
		\caption{The effect of magnetic field to the mass--radius relation, for zero field, Chandrasekhar's results (black squares), $(B_{\rm s},\,B_{0})=(10^{9},\,10^{14})$\,G\,(red circles), $(10^{9},\,10^{13})$\,G\,(blue upwards triangles), $(10^{9},\,10^{12})$\,G\,(sky diamonds), $(10^{7}$,\,$10^{14})$\,G\,(yellow stars) and $(10^{7},\,10^{12})$\,G\,(magenta downward triangles).}
		\label{fig:2}
	\end{figure}

	\begin{figure}
		\includegraphics[width =\columnwidth]{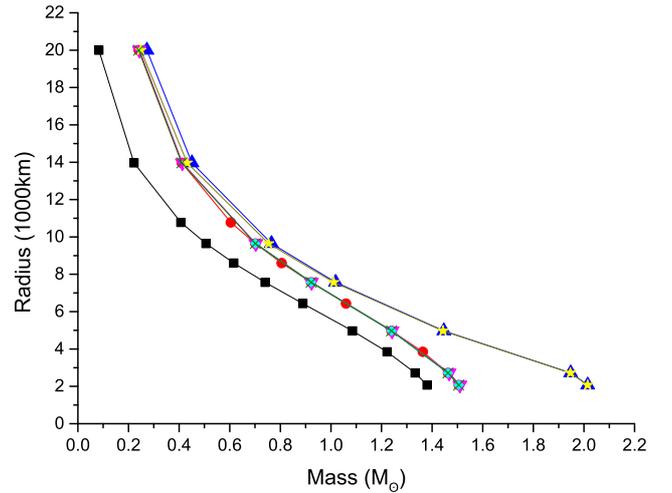}
		\caption{The effect of magnetic field and luminosity on the mass--radius relation, for Chandrasekhar's
			results, without thermal and field contributions (black squares), and $dT/dr\neq 0$ throughout for 
			$(B_{s},\,B_{0})=(0,\,0)\,$G\,(red circles), $(10^{9},\,10^{14})$\,G\,(blue upward triangles), $(10^{9},\,10^{13})$\,G\,(magenta downward triangles), $(10^{9},\,10^{12})$\,G\,(sky diamonds), $(10^{7},\,10^{14})$\,G\,(yellow stars) and $(10^{7},\,10^{12})$\,G\,(green cross). Except for Chandrasekhar's results, the luminosity is chosen to be $10^{-4}\,L_\odot$.}
		\label{fig:3}
	\end{figure}
	
	Table \ref{Table1} lists the central and interface temperatures and densities for some white dwarf 
	solutions of various masses and radii. It can be inferred that, for a particular radius of the 
	white dwarf, increasing the luminosity increases the effective thermal energy or temperature, 
	leading to a larger size of the surface layer or envelope. This in turn decreases the interface radius 
	in order to hold the excess thermal energy.
	Note that including the radiative transport of energy increases the central density and so also the 
	mass, compared to Chandrasekhar's results (\citealt{Chandra}). However, for smaller radii, when we 
	make the temperature constant below the interface radius, condition two above, the increase 
	in central density is not significant so that the Chandrasekhar-limit is preserved. This is also because the 
	surface layer is smaller for smaller white dwarfs because they are more dominated by electron degenerate matter.
	
	\begin{figure}
		\includegraphics[width =\columnwidth]{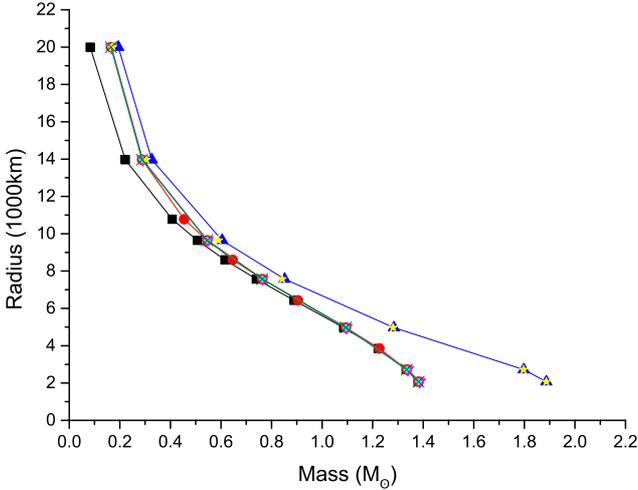}
		\caption{The effect of magnetic field and luminosity on the mass--radius relation, for Chandrasekhar's 
			results, without thermal and field contributions (black squares), and for $dT/dr=0$ from the interface to 
			the centre with $(B_{s},\,B_{0})=(0,\,0)$\,G\,(red circles), $(10^{9},\,10^{14})$\,G\,(blue upward triangles), $(10^{9},\,10^{13})$\,G\,(magenta downward triangles), $(10^{9},\,10^{12})$\,G\,(sky diamonds), $(10^{7},\,10^{14})$\,G\,(yellow stars) and $(10^{7},\,10^{12})$\,G\,(green cross). Except for Chandrasekhar's results, the luminosity is chosen to be $10^{-4}\,L_\odot$.}
		\label{fig:4}
	\end{figure}
	
	\begin{table*}
		\centering  
		\caption{The effect of luminosity on the mass--radius relation where $\rho_{\rm c}$, $T_{\rm c}$ and $M^{'}$ are respectively 
			central density, central temperature and mass, when $dT/dr=0$ below envelope, $R_{\star},\rho_{\star}$ and $T_{\star}$ are interface radius, density 
			and temperature respectively. $M$ is the mass for $dT/dr\neq 0$ throughout.}  
		\label{Table1}
		\begin{tabular}{||c c c c c c c c c||}
			\hline
		
			$R/1000\,$km &$L/L_{\odot}$ &$\rho_{\rm c}/10^{6}\,{\rm g}\,{\rm cm}^{-3}$ &$T_{\rm c}/10^{6}\,\rm K$ & $R_{\star}/1000\,$km &$\rho_{\star}/10^{6}\,\rm g\,cm^{-3}$ &$T_{\star}/10^{8}\,\rm K$ & $M^{'}/M_{\odot}$ & $M/M_{\odot}$\\
			\hline\hline
			20 & $10^{-4}$ & 0.0983 & 8.34 & 16.55 & 0.0011 & 0.083 & 0.165 & 0.24
			\\
			
			& $10^{-2}$ & 0.212 & 27.39 & 13.99 & 0.0068 & 0.274 & 0.265 & 0.37\\
			\hline
			9.645 & $10^{-4}$ & 2.36 & 5.92 & 9.442 & 0.00068 & 0.059 & 0.545 & 0.7
			\\
			
			& $10^{-2}$ & 3.15 & 21.29 & 9.03 & 0.0047 & 0.213 & 0.615 & 0.86\\
			\hline
			4.968 & $10^{-4}$ & 55.35 & 4.85 & 4.95 & 0.0005 & 0.0485 & 1.09 & 1.24
			\\
			
			& $10^{-2}$ & 60.29 & 17.97 & 4.89 & 0.0036 & 0.18 & 1.11 & 1.37\\
			\hline
			2.068 & $10^{-4}$ & 1962 & 4.54 & 2.0651 & 0.00046 & 0.0454 & 1.381 & 1.5
			\\
			
			& $10^{-2}$ & 2019 & 16.89 & 2.0572 & 0.0034 & 0.169 & 1.385 & 1.61\\
			[1ex]
			\hline
			
		\end{tabular}

	\end{table*}

	\begin{table*}
		\centering
		\caption{The effect of magnetic field for a particular luminosity ($10^{-4}L_{\odot}$) on the mass--radius 
			relation where $\rho_{\rm c}$, $T_{\rm c}$ and $M^{'}$ are respectively central density, central
			temperature and mass, when $dT/dr=0$ below interface, $R_{\star},\rho_{\star}$ and $T_{\star}$ are interface radius, density and temperature respectively. $M$ 
			is the mass for $\frac{dT}{dr}\neq 0$ throughout.}
		\label{Table2}
		\begin{tabular}{||c c c c c c c c c||}
			\hline
			
			$R/1000\,$km &$(B_{\rm s},B_{0})$/G &$\rho_{\rm c}/10^{6}\,{\rm g}\,{\rm cm}^{-3}$ &$T_{\rm c}/10^{6}\,\rm K$ & $R_{\star}/1000\,$km &$\rho_{\star}/10^{6}\,\rm g\,cm^{-3}$ &$T_{\star}/10^{8}\,\rm K$ & $M^{'}/M_{\odot}$ & $M/M_{\odot}$\\
			\hline\hline
			20 & $(10^{9},10^{14})$ & 0.136 & 8.04 & 15.93 & 0.00108 & 0.0804 & 0.194 & 0.27
			\\
			& $(10^{9},10^{13})$ & 0.102 & 8.3 & 16.46 & 0.00114 & 0.083 & 0.168 & 0.24
			\\
			& $(10^{7},10^{14})$ & 0.1 & 8.27 & 16.64 & 0.0011 & 0.0827 & 0.171 & 0.25
			\\
			\hline
			9.645 & $(10^{9},10^{14})$ & 2.68 & 5.8 & 9.346 & 0.00066 & 0.058 & 0.604 & 0.76
			\\
			& $(10^{9},10^{13})$ & 2.38 & 5.92 & 9.429 & 0.000683 & 0.059 & 0.547 & 0.70
			\\
			& $(10^{7},10^{14})$ & 2.5 & 5.79 & 9.458 & 0.00066 & 0.0579 & 0.59 & 0.75
			\\
			\hline
			4.968 & $(10^{9},10^{14})$ & 59.17 & 4.7 & 4.9342 & 0.00049 & 0.047 & 1.284 & 1.44
			\\
			& $(10^{9},10^{13})$ & 55.483 & 4.85 & 4.944 & 0.00051 & 0.0485 & 1.096 & 1.24
			\\
			& $(10^{7},10^{14})$ & 58.26 & 4.64 & 4.949 & 0.000475 & 0.0465 & 1.28 & 1.44
			\\
			\hline
			2.068 & $(10^{9},10^{14})$ & 2502.5 & 4.23 & 2.0642 & 0.00043 & 0.0423 & 1.887 & 2.01
			\\
			& $(10^{9},10^{13})$ & 1966.5 & 4.54 & 2.06486 & 0.000462 & 0.0454 & 1.39 & 1.51
			\\
			& $(10^{7},10^{14})$ & 2490.3 & 4.16 & 2.066 & 0.000414 & 0.0416 & 1.887 & 2.01
			\\
			[1ex]
			\hline
			
		\end{tabular}

	\end{table*}
	
	\section{Effect of Magnetic Field}
	In order to understand how magnetic field affects white dwarf structure, we first investigate its effect on the magnetostatic equilibrium equation 
	(\ref{eq:magnetostatic}) and mass conservation equation (\ref{eq:mass conservation}) 
	with $P_{\rm ideal}=0$, without the photon diffusion equation (\ref{eq:photon diffusion}). 
	It is seen from Fig. \ref{fig:2} that $B_{0}$ plays the crucial role and it has to be large enough 
	for any significant change in the mass--radius relation. Hence, except for the cases with 
	$(B_{\rm s},\,B_{0})=(10^{9},\,10^{14})$\,G\,(red circles) and $(10^{7},\,10^{14})$\,G\,(yellow stars), all the results practically
	overlap.

	Now if we include the photon-diffusion equation (\ref{eq:photon diffusion}) in the computation, 
	Figs \ref{fig:3} and \ref{fig:4} show how the magnetic field, along with luminosity, affects the mass--radius 
	relation for conditions one and two above. Only strong central magnetic fields significantly affect the mass of a white dwarf for a given radius. The mass--radius relations with central field $\sim 10^{14}$ G merge at high mass 
(central density) regime exceeding Chandrasekhar's limit, as shown by the lines with blue upward triangles and yellow stars in Figs \ref{fig:3} and \ref{fig:4}. Moreover, all other mass--radius relations of lower central fields also overlap with each other independent of magnetic fields, particularly
in the high mass (central density) regime, restricted by Chandrasekhar's limit. However, the mass--radius curve with $dT/dr=0$ throughout (without contribution
of photon diffusion) and without magnetic field, i.e. Chandrasekhar's result, shown by the line with black squares is
uniquely separated from other curves, as shown in Fig. \ref{fig:3}. This confirms that temperature gradient and magnetic fields
both have distinct effects in white dwarfs. However, Fig. \ref{fig:4} shows that photon diffusion with isothermal core 
($dT/dr=0$ below interface), with lower central magnetic fields $<10^{14}$ G, only affects the mass--radius relation at the low mass (density) regime slightly, leading to 
increasing radius for a given mass (see lines with red circles, magenta downward triangles, sky diamonds and green cross). 
In the high mass (density) regime, all low magnetic field mass--radius curves practically overlap along with that of Chandrasekhar.

	Notice, the central density of highly magnetized white dwarfs is higher than 
	that of the non-magnetized ones 
	in order to compensate for the additional magnetic pressure. This is because the contribution of the magnetic field 
	to the total pressure is larger than that to total density. 
	Interestingly, white dwarfs with a central field of about $10^{14}$\,G are significantly super-Chandrasekhar 
	with masses exceeding $2\,M_\odot$ at high central densities.
	
	Table \ref{Table2} shows how the presence of a magnetic field increases the central density of 
	white dwarfs and so increases their mass. Intuitively, we know that the presence of magnetic field decreases 
	the interface radius because it affects $\frac{dT}{dr}$ directly, similarly to the effect of luminosity on the system. 
	Therefore, higher magnetic fields have larger effects. Nevertheless, it is crucial to know that the 
	effective strength of the magnetic field is defined by both its $B_{\rm s}$ and $B_{0}$. 
	If we decrease $B_{0}$ for a particular $B_{\rm s}$, the results become closer to the non-magnetic case. 
	But for a high $B_{0}$, decreasing $B_{\rm s}$ also leads to a larger interface radius than the 
	non-magnetic case, as seen in Table~\ref{Table2} when compared with the respective cases in Table~\ref{Table1}.

	\begin{table*}
		\centering
		
		\caption{Mass and central density of white dwarf from conservation of magnetic and thermal energies for different field profiles and 
			radii. Here $M_{\rm org}$ and $\rho_{\rm c}^{\rm org}$ denote the mass and central density of the white dwarf in the absence of magnetic field 
with $L=10^{-4}~L_\odot$. Also, the central densities here are only for the cases With Interface. }
		\label{Table3}
		\begin{tabular}{||c c c c c c c c c c||} 
			\hline
			&  &  Without & Interface && With & Interface& && \\ [0.5ex] 
			\hline

			$R/1000\,$km & $(B_{\rm s},B_{0})$/G & $L/L_{\odot}$& $M/M_{\odot}$ & $M_{\rm org}\,/M_{\odot}$ &$L/L_{\odot}$& $M/M_{\odot}$ 
			& $M_{\rm org}\,/M_{\odot}$ & $\rho_{\rm c}/10^{6}\,\rm g\,cm^{-3}$ & $\rho_{\rm c}^{\rm org}/10^{6}\,\rm g\,cm^{-3}$\\
			\hline\hline
			20 & $(10^{9},10^{14})$ & $2\times 10^{-5}$ & 0.244 & 0.24 & $8\times 10^{-6}$ & 0.166 & 0.165 & 0.106 & 0.0983
			\\
			\hline
			20 & $(10^{7},10^{14})$ & $7\times 10^{-5}$ & 0.242 & 0.24 & $7\times 10^{-5}$ & 0.166 & 0.165 &0.095 & 0.0983\\
			\hline
			13.965 & $(10^{9},10^{14})$ & $2\times 10^{-5}$ & 0.416 & 0.411 & $10^{-6}$ & 0.287 & 0.288 & 0.3934 & 0.4095\\
			\hline
			13.965 & $(10^{7},10^{14})$ & $5\times10^{-5}$ & 0.415 & 0.411 & $3\times10^{-5}$ & 0.289 & 0.288 & 0.3854 & 0.4095\\
			\hline
			9.645 & $(10^{9},10^{14})$ & $5\times 10^{-6}$ & 0.701 & 0.701 & $10^{-12}$ & 0.566 & 0.545 & 2.29 & 2.36 \\
			\hline
			9.645 & $(10^{7},10^{14})$ & $10^{-5}$ & 0.702 & 0.701 & $10^{-12}$ & 0.551 & 0.545 & 2.121 & 2.36\\
			\hline
			7.57 & $(10^{9},10^{14})$ & $5\times 10^{-7}$  & 0.923 & 0.922 & $10^{-12}$ & 0.829 & 0.763 & 7.873 & 7.738\\
			\hline
			7.57 & $(10^{7},10^{14})$ & $10^{-6}$ & 0.924 & 0.922 & $10^{-12}$ & 0.818 & 0.763 & 7.7563 & 7.738\\
			\hline
			4.968 & $(10^{9},10^{14})$ &  $10^{-16}$ & 1.278 & 1.239 & $10^{-16}$ & 1.273 & 1.094 & 57.19 & 55.35\\
			\hline
			4.968 & $(10^{7},10^{14})$ & $10^{-12}$ & 1.285 & 1.239 & $10^{-12}$ & 1.270 & 1.094 & 56.35 & 55.35\\
			\hline
			2.7215 & $(10^{9},10^{14})$ & $10^{-12}$ & 1.807 & 1.463 & $10^{-16}$ & 1.794 & 1.335 & 720.89 & 686.96\\
			\hline
			2.7215 & $(10^{7},10^{14})$ & $10^{-12}$ & 1.807 & 1.463 & $10^{-12}$ & 1.793 & 1.335 & 718.62 & 686.96\\
			\hline
			2.068 & $(10^{9},10^{14})$ & $10^{-20}$ & 1.886 & 1.505 & $10^{-16}$ & 1.886 & 1.381 & 2473.8 & 1962
			\\
			\hline
			2.068 & $(10^{7},10^{14})$ & $10^{-20}$ & 1.886 & 1.505 & $10^{-16}$ & 1.886 & 1.381 & 2462.4 & 1962\\ [1ex]
			\hline
			
		\end{tabular}

	\end{table*}

	\section{Suppression of luminosity due to magnetic field}
	
	From the conservation of energy, in order to assure stability, the increase in the magnetic energy in 
	the white dwarf has to be compensated for by a decrease in thermal energy for a given gravitational energy.
	This in turn decreases the measure of luminosity and, hence, the temperature. 
This is particularly the case in lower density higher radius white dwarfs, when the thermal, magnetic and gravitational energies 
are comparable, i.e., when the ideal gas and magnetic pressures are comparable and the degenerate gas pressure 
inside white dwarfs is not too high compared to them.
From Figs \ref{fig:2}, \ref{fig:3} and
	\ref{fig:4}, 
	we have noticed that strong central magnetic fields corresponding to $B_{0} = 10^{14}$\,G show 
	a deviation from the non-magnetized mass--radius relation. We explore these cases further
and the results are listed in Table \ref{Table3}.
	We find that for larger radius lower density white dwarfs, a slight decrease in luminosity 
	leads to the mass and central density of the magnetized white dwarf comparable to its non-magnetic counterpart. This is still 
	in the observable range. 

\begin{figure}
		\includegraphics[width = \columnwidth]{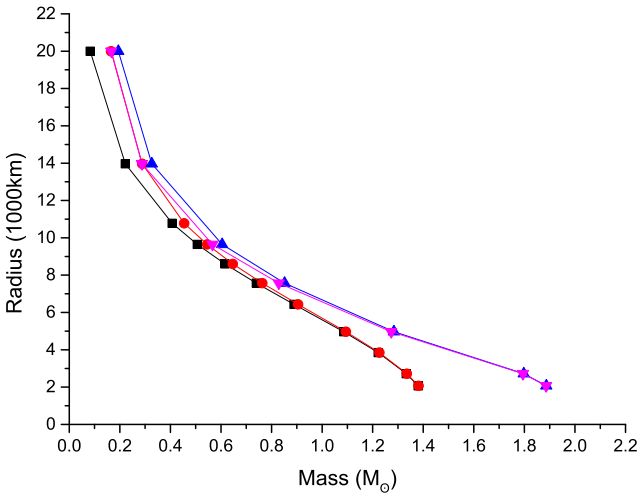}
		\caption{The effect of magnetic field optimizing luminosity to 
attempt to match with Chandrasekhar's mass--radius 
			relation (black squares), for $(B_{\rm s},\,B_{0})=(10^{9},\,10^{14})$\,G\,(magenta downward triangles), while
			the lines with red circles and blue upward triangles represent the results with $L=10^{-4}\,L_\odot$ for $(B_{\rm s},\,B_{0})=(0,\,0)$\,G and 
			$(10^{9},\,10^{14})$\,G. All cases correspond to $\frac{dT}{dr}=0$ below the interface radius. 
			See Table~\ref{Table3} for specific luminosities.}
		\label{fig:5}
	\end{figure}

However, the situation differs for higher density smaller radius white dwarfs, where
the thermal pressure inside turns out to be significantly sub-dominant, hence decreasing thermal energy does
not affect its mass. Moreover, as the magnetic pressure contributes to magnetostatic balance with increasing
mass for a radius same as the non-magnetic white dwarf, its central density increases compared to its non-magnetic counterpart in order to satisfy mass equation. See Table \ref{Table3}.
Therefore, for smaller radius higher density white dwarfs, even if the luminosity decreases 
	considerably ($10^{-12}$ to $10^{-20}\,L_{\odot}$), the mass of the magnetized white dwarf remains
	quite high compared to its non-magnetic counterpart. This leads to an extended branch in the mass--radius
relation which is absent in the non-magnetic counterpart. This is seen in Fig. \ref{fig:5}. 
	This implies that there can be white dwarfs with high magnetic field strengths (and high mass) but these are too faint to detect.
	
	\section{Discussion}
	
	The SDSS data show that there are a large number of large radius and small mass white dwarfs 
	whose masses are high
	compared to Chandrasekhar's mass--radius relation. Nevertheless,
	\cite{Rueda} showed that the finite temperature electron degenerate EoS at higher temperatures
	could explain these observations. We also obtain similar results for no and weak magnetic fields by 
	appropriately introducing the photon diffusion equation (\ref{eq:photon diffusion}). Moreover, our solutions 
	self-consistently define the interface between the interior degenerate core and outer envelope with ideal 
	gas layers. 
	
	In addition, we obtain super-Chandrasekhar non-magnetic white dwarfs for a 
	constant luminosity of, say,  
	$10^{-2}\,L_{\odot}$ hypothesizing $dT/dr\neq 0$ throughout. 
	This particularly shows the effects of very high temperatures, which arise 
	because of the nonzero $dT/dr$ 
	throughout and high surface temperature, on small radius white dwarfs.
	Further, in the presence of weak magnetic fields, we obtain the same mass--radius relations as for the 
	non-magnetic cases when  $dT/dr=0$ below the interface. However in cases of stronger magnetic fields, 
	whether $dT/dr=0$ below the interface or not, we obtain super-Chandrasekhar masses. Thus our results, 
	which depend on the photon diffusion equation,
	suggest the existence of super-Chandrasekhar white dwarfs for three different cases, (1) 
	constant higher luminosity (about $10^{-2}\,L_{\odot}$) with nonzero $dT/dr$ throughout, 
	(2) strong magnetic field with non-zero $\frac{dT}{dr}$ throughout, and (3) strong magnetic field
	with $\frac{dT}{dr}=0$ below the interface radius. While the two former situations may be implausible because
	then nonmagnetic and nonrotating white dwarfs might have super-Chandrasekhar mass, the
	last is quite plausible. These could very well be the candidates for
	super-Chandrasekhar progenitors of overluminous Type Ia supernovae. 
	
	Another interesting result is that, for the white dwarfs with smaller radii and stronger magnetic fields, 
	the decrease in luminosity does not lead to the merging of the mass--radius 
	relation of white dwarfs with that of their nonmagnetic counterparts,
	leaving super-Chandrasekhar white dwarfs separate in the mass--radius relation.
	Hence, there can be super-Chandrasekhar magnetic white dwarfs which are so dim or have very low luminosity 
	(as little as $10^{-20}\,L_{\odot}$), which we have missed from detection. But for larger radii, depending on their 
	magnetic field strengths, we see the mass--radius relation merging with the
	nonmagnetic counterpart. Hence, 
	with improved observational techniques and instruments, these still hold a chance to be observed because their 
	luminosity is of the order of $10^{-5}$ to $10^{-8}\,L_{\odot}$. 
	
	There are a few important points to be stressed. In our case, the EoS for the electron degenerate pressure 
	is temperature independent and it becomes very important to check for its validity when we are 
	solving for $\frac{dT}{dr}$. In order to validate the choice, the Fermi energy $E_F$ has to be much greater 
	than the thermal energy of the matter. For all the cases considered here, this condition has always been 
	satisfied, at least in the inner degenerate core. It is observed that $E_{F}/kT \approx 5$ at the interface, 
	and it increases below the interface.
	
	We have attempted to explore the idea of including the energy conservation equation 
	$\frac{d L(r)}{d r}=4 \pi r^{2} \rho(r) \epsilon(r)$, where $\epsilon(r)$ is the power produced per 
	unit mass of stellar material in the system, in place of constant luminosity. However, because white 
	dwarfs do not have a consistent energy production mechanism, our model equation does not perform 
	adequately compared to our assumption of constant luminosity. 
Further, at high temperature, energy losses due to neutrino emission may
turn out to be important (depending on the temperature and its variation),
which we also have omitted from the present work. Moreover, we have
not included possible convection in our computation, assuming that strong 
field inhibits its effect. However, convection may transfer significant 
energy (\citealt{2014Natur.515...88V}) in nonmagnetic and weakly magnetized 
white dwarfs which we keep considering for comparison, particularly for 
$dT/dr\neq 0$ throughout. In future, all of these physics will be 
appropriately considered in a more sophisticated model framework.
	
	In addition to this, we have used a Newtonian framework throughout our work.
	This is practically a valid
	assumption for the present purpose of white dwarfs. We eventually should explore the Tolman-Oppenheimer-Volkoff 
	equation (\citealt{Tol}) along with temperature gradient and
necessary related physics. Nonetheless, in the case of white dwarfs, $2GM/c^{2}R$ is quite small, and this favours the choice of a Newtonian framework. 
	
	The work may be further extended for higher magnetic fields, where Landau quantization becomes important. Nonetheless, this work, in itself, is quite unique, because we could define an interface between the degenerate core and outer ideal gas envelope self-consistently and obtain mass--radius and luminosity relationships including the radiative transport equation in the 
	magnetized and non-magnetized cases, which are in accordance with observations. Moreover, we have 
	obtained a unique result that, in the presence of strong magnetic fields, small radius white dwarfs can always 
	be undetected because they may be too dim to detect, even if they are on the similar mass--radius track
	as their nonmagnetic counterparts. 
	
	\section{Conclusion}
	Observations of the light curves of over-luminous peculiar type 1a supernovae imply the presence of super-Chandrasekhar white dwarfs but there has been no direct observation of such a WD. Moreover, many theories have been put forward to explain the Chandrasekhar mass limit violation. These include the white dwarfs being highly magnetized and rotating. We argue that highly magnetized white dwarfs' luminosities can be suppressed to the extent that they are practically hidden, and cannot be detected. In the future, through gravitational wave detectors such as LISA, eLISA and a few others, their presence may be detected (\citealt{kalitagw}). 
	
	Most of the studies of white dwarfs, so far, assumed beforehand the existence of an interface between the two pressure components, electron degenerate pressure and ideal gas pressure. We have shown that the interface can be self-consistently found, for magnetic or non-magnetic white dwarfs. 
	
	\section*{Acknowledgements}
	AG and BM would like to thank Mukul Bhattacharya of the University of Texas, Austin for discussion at the 
	initial phase of the work. This work was partly supported by a project of 
	Department of Science and Technology with Grant No.DSTO/PPH/BMP/1946 (EMR/2017/001226). CAT thanks Churchill College for his fellowship.
	
	
	
	
	\bibliographystyle{mnras}
	\bibliography{references1} 

	
	
	\appendix
	
	\section{Representative sample case study}
	
	Here we outline the process of obtaining the interface radius for a particular case. 
	The following example is for $R=4968$\,km white dwarf in the presence of magnetic field 
	($B_{\rm s},\,B_{0})=(10^{9},\,10^{14}$)\,G
	with luminosity $10^{-4}\,L_{\odot}$.
	
\begin{figure*}
		\centering
		\includegraphics[width=.45\linewidth]{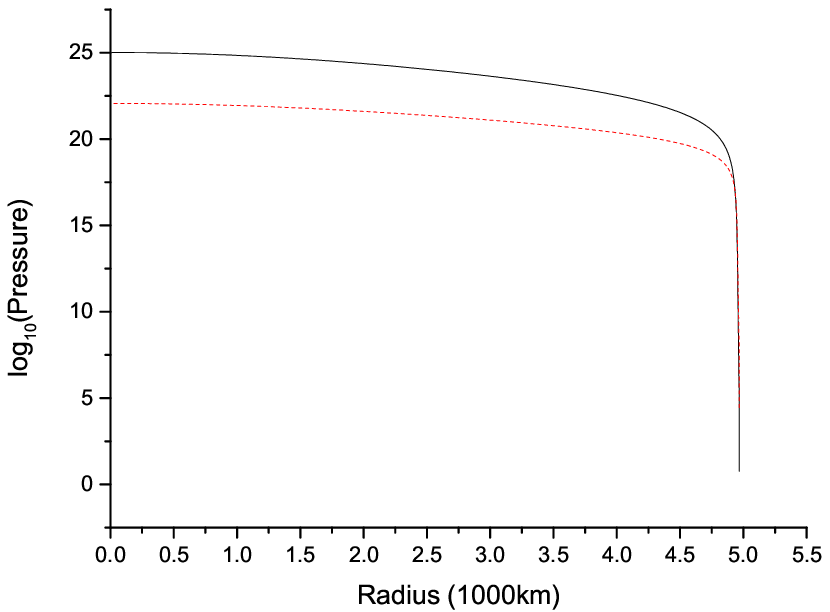}
		\includegraphics[width=.45\linewidth]{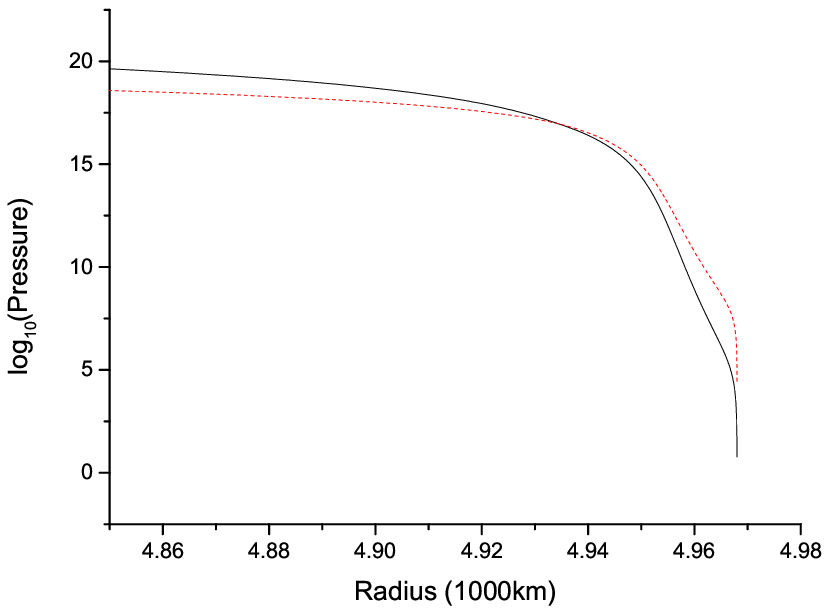}
		\caption{
			The variation of electron degenerate pressure (solid black line) and ideal gas pressure (dash red line) as functions of radius, when $dT/dr=0$ below interface.
			The right panel is the same as left panel, but zoomed in the outer region close to the surface. Note, the ideal gas pressure takes over the electron degenerate pressure around $4934$\,km.
		}
		\label{fig:9}
	\end{figure*}

\begin{figure*}
		\includegraphics{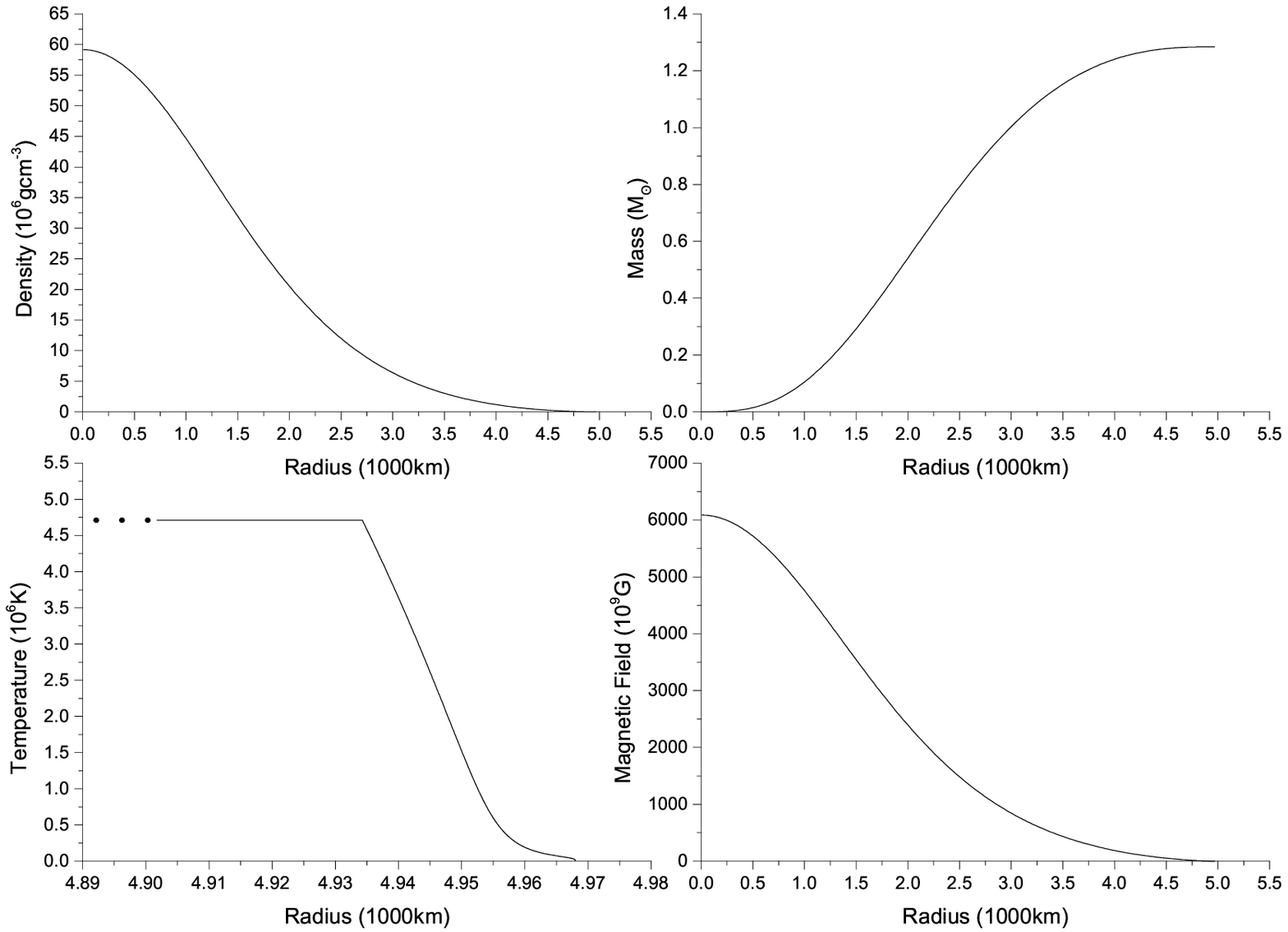}
		\caption{For a particular mass--radius, the variations of density (top left), mass (top right), temperature (bottom left) and magnetic field (bottom right), as
			functions of radius in a white dwarf. Note that the 
temperature varies only on the surface layer above $4934$\,km (interface) 
and remains constant from the interface to centre, part of which has been depicted.} 
		\label{fig:6}
	\end{figure*}

\begin{figure*}
		\includegraphics{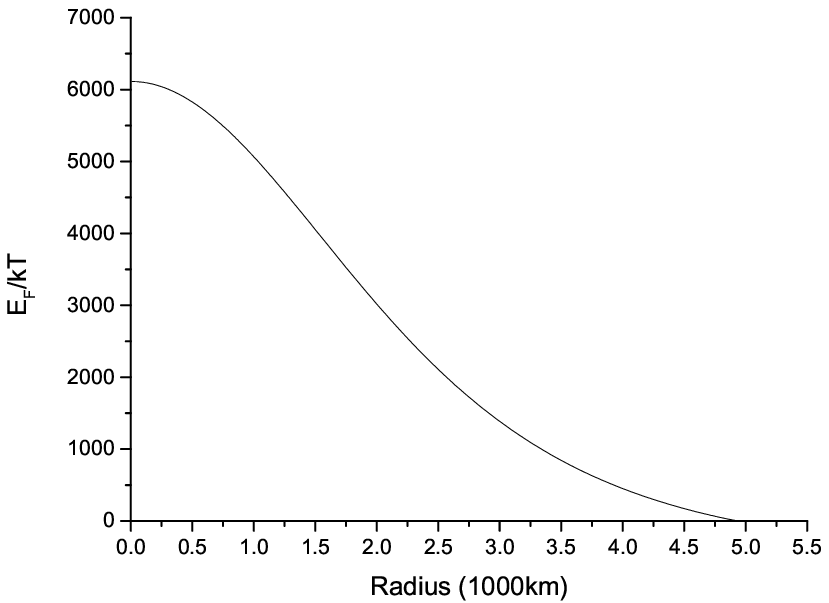}
		\caption{The variation of ratio of the Fermi energy to thermal energy as a function of radius.
		}
		\label{fig:8}
	\end{figure*}

	In Fig. \ref{fig:9}, we show the various pressure profiles and it is 
	seen that roughly at $4934$\,km, the ideal gas pressure 
	overpowers the electron degeneracy pressure. 
	Hence, this radius can be defined as the interface radius for this 
	case. Fig. \ref{fig:9} clearly confirms the 
	dominance of the electron degenerate matter in the core about three orders of magnitude over the ideal gas pressure.
	Further, applying $\frac{dT}{dr}=0$ below this interface radius, we obtain the mass for $R=4968$\,km 
	white dwarf, see Table \ref{Table2}. 
	Fig. \ref{fig:6} represents the profiles for density, mass, temperature and magnetic field as
	functions of radius. Density and magnetic field have roughly the same profiles because the strength of magnetic 
	field is density dependent. 
	Fig. \ref{fig:8} also confirms that Fermi energy is significantly dominant over thermal energy, assuring 
	the validity of zero-temperature EoS.
	
	
	
	\bsp	
	\label{lastpage}
\end{document}